\documentclass{article}

\usepackage{spconf}
\usepackage{amsmath}
\usepackage{graphicx}

\usepackage{amsfonts}
\usepackage[english]{babel}
\usepackage{booktabs}
\usepackage{csquotes}
\usepackage[T1]{fontenc}
\usepackage{mathtools}
\usepackage{microtype}
\usepackage{siunitx}
\usepackage{subcaption}
\usepackage{tabularx}
\usepackage{url}

\graphicspath{{figures/}}

\DeclareGraphicsExtensions{.pdf}

\newcommand{\Xcal}{\mathcal{X}}
\newcommand{\Zcal}{\mathcal{Z}}
\newcommand{\JSD}{\mathrm{JSD}}
\newcommand{\ce}{\ell_{\mathsf{ce}}}
\newcommand{\jce}{\ell_{\mathsf{jce}}}
\newcommand{\closs}{\ell_{\mathsf{sim}}}

\DeclarePairedDelimiterX{\infdivx}[2]{(}{)}{%
  #1\;\delimsize\|\;#2%
}
\newcommand{\KL}{\mathrm{KL}\infdivx}

\newcommand{\score}[2]{\(#1\% \scriptstyle{\pm #2}\)}
\newcommand{\scoreb}[2]{\(\mathbf{#1\% \scriptstyle{\pm #2}}\)}
\newcommand{\offset}[1]{\(#1\%\)}
\newcommand{\offsetb}[1]{\(\mathbf{#1\%}\)}

\title{Enhancing Audio Augmentation Methods with Consistency Learning}

\name{Turab Iqbal$^1$, Karim Helwani$^2$, Arvindh Krishnaswamy$^2$, Wenwu Wang$^1$}
\address{$^1$Centre for Vision, Speech and Signal Processing, University of Surrey, UK\\
         $^2$Amazon Web Services, Inc., Palo Alto, CA, USA\\
         \small{\tt{\{t.iqbal,w.wang\}@surrey.ac.uk, \{helwk,arvindhk\}@amazon.com}}}

\begin{document}

\maketitle

\begin{abstract}
Data augmentation is an inexpensive way to increase training data diversity
and is commonly achieved via transformations of existing data. For tasks such
as classification, there is a good case for learning representations of the
data that are invariant to such transformations, yet this is not explicitly
enforced by classification losses such as the cross-entropy loss. This paper
investigates the use of training objectives that explicitly impose this
consistency constraint and how it can impact downstream audio classification
tasks. In the context of deep convolutional neural networks in the supervised
setting, we show empirically that certain measures of consistency are not
implicitly captured by the cross-entropy loss and that incorporating such
measures into the loss function can improve the performance of audio
classification systems. Put another way, we demonstrate how existing
augmentation methods can further improve learning by enforcing consistency.
\end{abstract}

\begin{keywords}
Audio classification, data augmentation, consistency learning, neural networks
\end{keywords}

\section{Introduction}
\label{sec:intro}

For tasks such as audio classification, a \textit{de facto} practice when
training deep neural networks is to use data augmentation, as it is an
inexpensive way to increase the amount of training data. The most common
approach to data augmentation is to use transformations of existing training
data. Examples of such transformations for audio include time-frequency
masking, the addition of noise, pitch shifting, equalization, and adding
reverberations \cite{da_salamon, specaug_park, poconet_isik}. These
transformations are intended to preserve the semantics of the data, so that for
an instance \(x\) belonging to a class \(y\), a transformation \(x'=T(x)\)
should also map to \(y\). From the perspective of representation learning, it
is also desirable for the model's latent representation of the data to capture
the data's properties \cite{repr_bengio}. This means that the representation
should remain unchanged or only slightly changed under these transformations
too. By learning representations that behave in this way, tasks such as
classification can benefit in terms of improved robustness to nuisance factors
and better generalization performance \cite{repr_bengio, stability_zheng,
augmix_hendryks}.

The standard cross-entropy function does not enforce this invariance constraint
explicitly. That is, the trained model's representations of instances \(x\) and
\(x'\) may differ significantly. To impose consistency between similar
instances, there has been interest in incorporating suitable similarity
measures into the training objective -- either as a standalone loss or as an
additional loss term. In these contexts, they are sometimes referred to as
consistency losses \cite{augmix_hendryks, cutmix_ss_french} or as stability
losses \cite{stability_zheng}. Closely related to this are contrastive losses
and triplet losses \cite{facenet_schroff, simclr_chen, triplet_jansen,
triplet_turpault}, where the objective is to cluster instances that are similar
while also separating instances that are dissimilar. Unlike the cross-entropy
loss, consistency losses and their variants do not require ground truth
labels, and have thus been adopted in unsupervised and semi-supervised settings
\cite{cutmix_ss_french, triplet_jansen, triplet_turpault}. In the image domain,
their efficacy has also been demonstrated for supervised learning in terms of
improving robustness against distortions \cite{stability_zheng,
augmix_hendryks}.

In this paper, we investigate the use of consistency losses for audio
classification tasks in the supervised learning setting. We examine several
audio transformations that could be used for data augmentation, and impose
consistency in a suitable latent space of the model when using these
transformations. We are interested in whether enforcing consistency can
influence the learned representations of the neural network model in a
significant way, and, if so, whether this is beneficial for downstream audio
classification tasks. An affirmative outcome would give a new purpose to data
augmentation methods and further enhance their utility. To our knowledge, this
is the first study in this direction for audio classification.

More concretely, we propose using the Jensen-Shannon divergence as a
loss term to constrain the class distribution \(P(\hat{Y} \mid X)\) of
the neural network to not deviate greatly under certain
transformations. On the ESC-50 environmental sound dataset \cite{esc_piczak},
the proposed method is shown to bring notable improvements to existing
augmentation methods. By tracking the Jensen-Shannon divergence as training
progresses, regardless of whether it is minimized, claims about the consistency
of the model outputs are verified. We also discover that the cross-entropy loss
on its own can encourage consistency to some extent if the data pipeline is
modified to include \(x\) and its transformations in the same training
mini-batch -- a variation we call \textit{batched data augmentation}.

\subsection{Related Work}
\label{ssec:related}

In terms of using consistency learning, the closest analog to our work is the
AugMix algorithm \cite{augmix_hendryks}, where the authors also propose the
Jensen-Shannon divergence as a consistency loss. Another related work is from
Zheng et al. \cite{stability_zheng}, where they use the Kullback-Leibler
divergence for class distributions and the \(L_2\) distance for feature
embeddings. These works look at improving robustness for image recognition when
distortions are present, while our work is on general audio recognition. In
addition, they only use the transformations to minimize the consistency loss
and not the cross-entropy loss. We observed that the benefits of augmentation
are partially lost this way. \mbox{A consistency} learning framework was also
proposed by Xie et al. \cite{uda_xie} but for unsupervised learning of
non-audio tasks.

A similar paradigm is contrastive learning, where, using a similarity measure,
a margin is maintained between similar instances and dissimilar instances in an
unsupervised fashion. In the audio domain, contrastive learning has been
explored for unsupervised and semi-supervised learning \cite{triplet_jansen,
triplet_turpault, wav2vec_schneider}. Another related concept is virtual
adversarial training (VAT) \cite{vat_miyato, covat_kreyssig}, which also
promotes consistency but for adversarial perturbations of the data.

\section{Consistency Learning}
\label{sec:consistency}

We first develop the consistency learning framework that our proposed method is
based on. A neural network is a function \(f: \Xcal \to \Zcal\) composed of
nonlinear, differentiable functions, \(f_l\), such that \(f=f_L \circ \ldots
\circ f_1\). Each \(f_l\) corresponds to a layer in the neural network. Using a
learning algorithm, the parameters of \(f\) are optimized with respect to a
suitable training objective. For a classification task with \(K\) classes,
\(f(x)\) is a vector of \(K\) class probabilities from which the class that
\(x\) belongs to can be inferred. The objective, in this case, is to minimize
the classification error, or rather a surrogate of this error that is
differentiable. In the supervised setting, this surrogate is most commonly the
cross-entropy loss function, \(\ce\).

Each layer of \(f\) produces a latent representation of the data. For certain
architectures, including the convolutional neural network architectures popular
in image/audio classification, earlier layers tend to capture the low-level
properties of the data after training, while the later layers tend to capture
the high-level properties \cite{repr_bengio}. Therefore, when concerned with
the data's high-level properties, the output of the penultimate layer,
\(f_{L-1}\), or sometimes the output of the final layer, \(f_L\), is considered
to be the representation of interest. We will denote as \(G(x)\) such a
representation of \(x \in \Xcal\).

Since \(G(x)\) is intended to capture high-level features of \(x\), it should
be insensitive to small perturbations of \(x\), such that \(G(x) \approx
G(T(x))\) for any \(T(x)\) that preserves such features. In particular, this
property should hold for the transformations used in data augmentation. The
motivation for learning such representations is to improve downstream
classification tasks. However, the cross-entropy loss does not explicitly
enforce this property of consistency. To enforce consistency, we can define a
similarity measure, \(D(G(x), G(x'))\), that is to be minimized when
\(x'=T(x)\) is a perturbation of \(x\). To do this, we add the similarity
measure as a loss term, giving:
\begin{align}
  \label{eq:loss}
  \ell'(x,x',y) &\coloneqq \ce(f(x),y) + \lambda \closs(x,x'), \\
  \closs(x,x') &\coloneqq D(G(x),G(x')),
\end{align}
where \(y\) is the ground truth label and \(\lambda\) determines the strength
of the new consistency loss term. Note that the cross-entropy loss can also be
minimized with respect to \(x'\), since \(x'\) belongs to class \(y\) by
design. Therefore, we modify \eqref{eq:loss} to give
\begin{align}
  \label{eq:loss2}
  \ell(x,x',y) &\coloneqq \jce(x,x',y) + \lambda \closs(x,x'), \\
  \label{eq:jce}
  \jce(x,x',y) &\coloneqq \frac{1}{2}[\ce(f(x),y) + \ce(f(x'),y)].
\end{align}
The training objective is then to minimize \eqref{eq:loss2} rather than the
cross-entropy loss on its own. This formulation can also be generalized to
handle multiple transformations \(x^1,\ldots,x^n\) of \(x\) provided that the
measure \(D\) can accommodate them.

\section{Proposed Method}
\label{sec:method}

Given the framework based on \eqref{eq:loss2}, the main considerations for
implementing consistency learning are the choices of the transformations and
the exact form of \(D\). These concerns are addressed in the following
subsections.

\subsection{Transformations}
\label{ssec:transforms}

This paper considers three types of transformations, which are:
\begin{itemize}
  \itemsep -0.15em
  \item \textbf{Pitch shifting:} The pitch of the audio clip is shifted to be
    higher or lower without affecting the clip's duration. The pitch is
    randomly shifted by \(l\) semitones, where \(l \in
    \{-2.5,-2,-1.5,-1,-0.5,0.5,1,1.5,2,2.5\}\).
  \item \textbf{Reverberations:} Reverberations are added to the audio by
    convolving the waveform with a randomly-generated room impulse response
    (RIR). The RIRs were set to have an RT60 between \SI{200}{\ms} and
    \SI{1000}{\ms}. We generated \num{500} RIRs in advance and selected one at
    random each time a transformation needed to be applied.
  \item \textbf{Time-frequency masking:} Regions of the spectrogram are
    randomly masked out in an effort to encourage the neural network to
    correctly infer the class despite the missing information. This is done
    in the same way as the SpecAugment algorithm \cite{specaug_park}, such that
    the number, size, and position of the regions is random. This is the only
    transformation that is applied to the spectrogram rather than the audio
    waveform directly.
\end{itemize}

Each type of transformation can produce several variations. A specific
variation is selected randomly each time an instance \(x\) needs to be
transformed. We apply \textit{two} transformations to each instance \(x\),
giving the triplet \((x, x^1, x^2)\). The consistency learning objective is
then to ensure \(G(x)\), \(G(x^1)\), and \(G(x^2)\) do not diverge from each
other. We found that applying two transformations rather than one improved the
performance of the trained model by a significant margin. It also allows for
greater diversity, because \(x^1\) and \(x^2\) can be generated using different
types of transformations.

Since the training instances must be processed in triplets to use our method,
the mini-batches used for training must be of size \(3N\), where \(N\) is the
number of original instances. As an ablation study, we also examine the case
in which this batch arrangement is used \textit{without} the consistency loss
term, giving \(\jce(x,x^1,x^2,y)\) as the loss instead. We refer to this as
\textit{batched data augmentation (BDA)}.

\subsection{Similarity measure}
\label{ssec:measure}

The representation \(G(x)\) we adopt in this paper is the output of the final
layer of the neural network, i.e. \(G(x) \coloneqq f(x)\). This means \(G(x)\)
represents a class probability distribution, \(P(\hat{Y} \mid X=x)\), which
is the model's estimate of the true class probability distribution, \(P(Y \mid
X=x)\). This choice of \(G(x)\) has high interpretability, since it is a vector
of probabilities associated with the target classes. Other latent
representations typically demand some form of metric learning before they can
be endowed with a similarity measure and interpreted \cite{simclr_chen}. Since
\(G(x)\) is a probability distribution, familiar probability distribution
divergences can be used directly.

We propose to use the Jensen-Shannon (JS) divergence as the similarity measure
\(D\). Given that we wish to measure the similarity between three distributions
-- \(P_x\), \(P_{x^1}\), and \(P_{x^2}\), where \(P_x \coloneqq
P(\hat{Y} \mid X=x)\) -- the JS divergence is defined as
\begin{equation}
\begin{split}
  \label{eq:jsd}
  \JSD(P_x,P_{x^1},P_{x^2}) \coloneqq \frac{1}{3}&[\KL{P_x}{M}\\
                                                 &+\KL{P_{x^1}}{M}\\
                                                 &+\KL{P_{x^2}}{M}],
\end{split}
\end{equation}
where \(M \coloneqq \frac{1}{3}(P_x + P_{x^1} + P_{x^2})\) and \(\KL{P}{Q}\) is
the Kullback-Leibler (KL) divergence from \(Q\) to \(P\). The primary reason for
using the JS divergence is that it can handle an arbitrary number of
distributions, while other divergences such as the KL divergence are defined
for two distributions only.

\subsection{Learning dynamics of consistency loss}
\label{ssec:dynamics}

Imposing consistency is arguably less meaningful when the neural network
outputs incorrect predictions. In these cases, the consistency loss may
negatively affect the learning process. To avoid this, we propose to linearly
increase the weight \(\lambda\) from zero to a fixed value for the first \(m\)
epochs. The rationale is that mispredictions are common at the beginning of
training but less likely as training progresses. Our experiments showed a
measurable improvement when using this heuristic.

\section{Experiments}
\label{sec:experiments}

In this section, we present experiments to evaluate our method. Our
intention is to compare the performance of standard data augmentation methods
to the proposed method, which uses the same audio transformations but with a
different data pipeline and a different loss function. The modified data
pipeline on its own corresponds to BDA (see Section \ref{ssec:transforms}),
which is also compared in our experiments. The models used for training are
convolutional neural networks (CNNs) with log-scaled mel spectrogram inputs.
These CNN models were evaluated on the ESC-50 environmental sound
classification dataset \cite{esc_piczak}.

\subsection{Dataset}
\label{ssec:datasets}

The ESC-50 dataset is comprised of \num{2000} audio recordings for
environmental audio classification. There are \num{50} sound classes, with
\num{40} recordings per class. Each recording is five seconds in duration and
is sampled at \SI{44.1}{\kHz}. The recordings are sourced from the Freesound
database\footnote{\url{https://freesound.org}} and are relatively free of
noise. The dataset creators split the dataset into five folds for the purpose
of cross-validation. To evaluate the systems, we use the given cross-validation
setup and report the accuracy, which is the percentage of correct predictions.

\subsection{Model architecture}
\label{ssec:model}

The neural network used in our experiments is a CNN based on the VGG
architecture \cite{vgg_simonyan}. The main differences are the use of batch
normalization \cite{batchnorm_ioffe}, global averaging pooling after the
convolutional layers, and only one fully-connected layer instead of three. The
model contains eight convolutional layers with the following number of output
feature maps: 64, 64, 128, 128, 256, 256, 512, 512. The inputs to the neural
network are mel spectrograms. To generate the mel spectrograms, we used a
short-time Fourier transform (STFT) with a window size of \num{1024} and a hop
length of \num{600}. \num{128} mel bins were used.

The models were trained using the AdamW optimization algorithm
\cite{adamw_loshchilov} for \num{72} epochs with a weight decay of \num{0.01}
and a learning rate of \num{0.0005}, which was decayed by \SI{10}{\%} after
every two epochs. The mini-batch size was set
to \num{120}. For our proposed method and the BDA method, this means there were
\num{40} original instances and \num{80} transformations per mini-batch.
Although some models converged sooner than \num{72} epochs, the
validation accuracy did not degrade with further training.

The consistency loss term of \eqref{eq:loss2} has one hyperparameter,
\(\lambda\), which is the weight. Following the discussion in Section
\ref{ssec:dynamics}, we initially set the weight to zero and linearly increased
it after each epoch until the \(m\)th epoch, at which point it remained at a
fixed value. In our experiments, \(m=10\) and the fixed value is \(\lambda=5\).
These hyperparameter values were selected using a validation set, though
we found that rigorous fine-tuning was not necessary (e.g. \(\lambda=7.5\) gave
similar results).

\subsection{Evaluated models}

For our experiments, we trained \num{13} types of models, with one of which
being the CNN without data augmentation applied. Four of the models apply
standard data augmentation, i.e. the training set is simply augmented and
instances are sampled from it as normal. They are \textit{Pitch-Shift},
\textit{Reverb}, \textit{TF-Masking}, and \textit{Combination}. As the names
imply, three of these apply just a single type of transformation (cf. Section
\ref{ssec:transforms}). \textit{Combination} applies either pitch-shifting or
reverberations randomly with equal probability. Complementing the
aforementioned four models are the BDA variations, which are suffixed with
\textit{-BDA} in the results table; and the variations using our consistency
learning method, which are suffixed with \textit{-CL}.

\subsection{Results}

\begin{table}[t]
  \caption{The experimental results for ESC-50. The average accuracy
           and standard error are stated along with the absolute improvement
           compared to using no data augmentation.}
  \label{t:results}
  \centering
  \begin{tabularx}{0.48\textwidth}{lXXXX}
    \toprule
    Model            & Accuracy             & Improvement \\
    \midrule
    No Augmentation  & \score{83.59}{0.15}  & - \\
    \midrule
    Pitch-Shift      & \score{84.35}{0.25}  & \offset{+0.76} \\
    Pitch-Shift-BDA  & \score{84.60}{0.31}  & \offset{+1.01} \\
    Pitch-Shift-CL   & \scoreb{85.48}{0.22} & \offsetb{+1.89} \\
    \midrule
    Reverb           & \score{83.76}{0.19}  & \offset{+0.17} \\
    Reverb-BDA       & \score{85.12}{0.07}  & \offset{+1.53} \\
    Reverb-CL        & \scoreb{85.58}{0.22} & \offsetb{+1.99} \\
    \midrule
    TF-Masking       & \score{83.69}{0.20}  & \offset{+0.10} \\
    TF-Masking-BDA   & \score{84.32}{0.29}  & \offset{+0.73} \\
    TF-Masking-CL    & \scoreb{85.03}{0.12} & \offsetb{+1.44} \\
    \midrule
    Combination      & \score{83.99}{0.26}  & \offset{+0.40} \\
    Combination-BDA  & \score{85.83}{0.22}  & \offset{+2.25} \\
    Combination-CL   & \scoreb{86.22}{0.12} & \offsetb{+2.63} \\
    \bottomrule
  \end{tabularx}
\end{table}

The results are presented in Table \ref{t:results}. The vanilla CNN achieves an
accuracy of \SI{83.59}{\%}, which matches results presented in the past for
such an architecture \cite{pann_kong}. The models implementing standard
augmentation improve the performance marginally, with an average accuracy
increase of \SI{0.36}{\%}. For batched data augmentation (BDA), sizable
improvements can be observed for all of the transformations (average
improvement of \SI{1.38}{\%}), including the \textit{Combination} variant, where
the improvement over the vanilla CNN is \SI{2.25}{\%}. Using the consistency
loss, the improvements are even greater, with an average accuracy increase of
\SI{1.99}{\%} and a maximum increase of \SI{2.63}{\%} when combining
transformations. Overall, these results show that consistency learning can
benefit audio classification and that BDA is also superior to standard
augmentation. It should be noted that using two transformations per instance
instead of one made a large difference in our experiments.

\newpage

\begin{figure}[t]
  \centering
  \begin{subfigure}[b]{0.98\columnwidth}
    \includegraphics[width=1\linewidth]{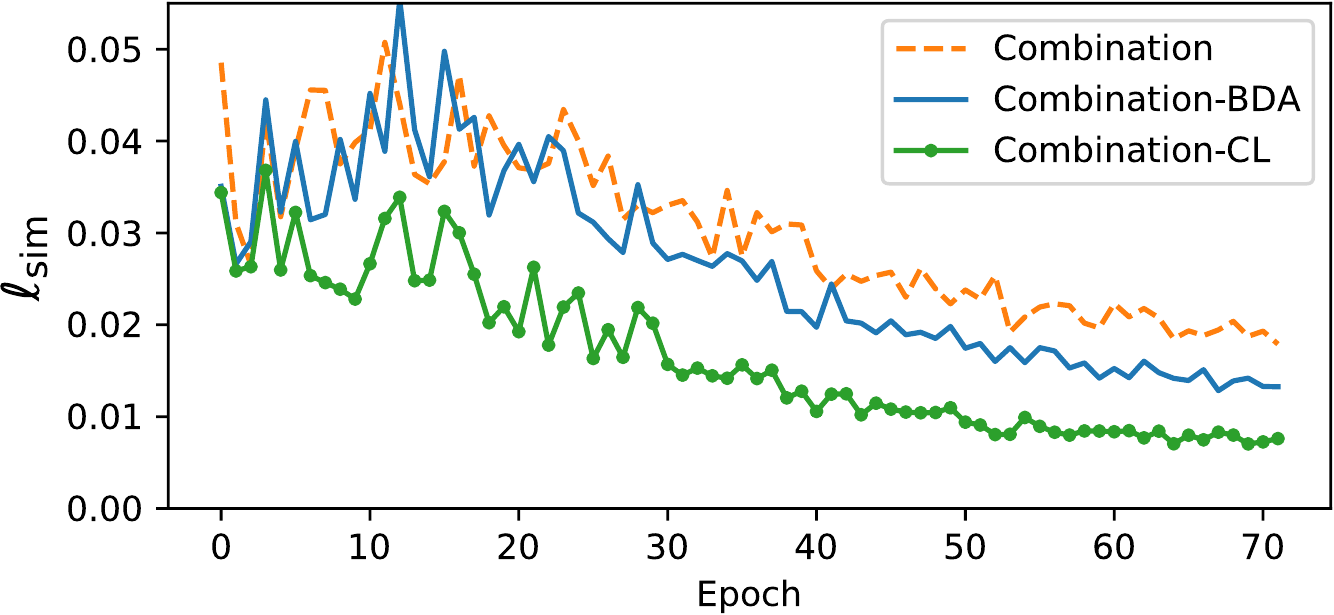}
    \caption{}
    \label{fig:js_loss_train} 
  \end{subfigure}

  \begin{subfigure}[b]{0.98\columnwidth}
    \includegraphics[width=1\linewidth]{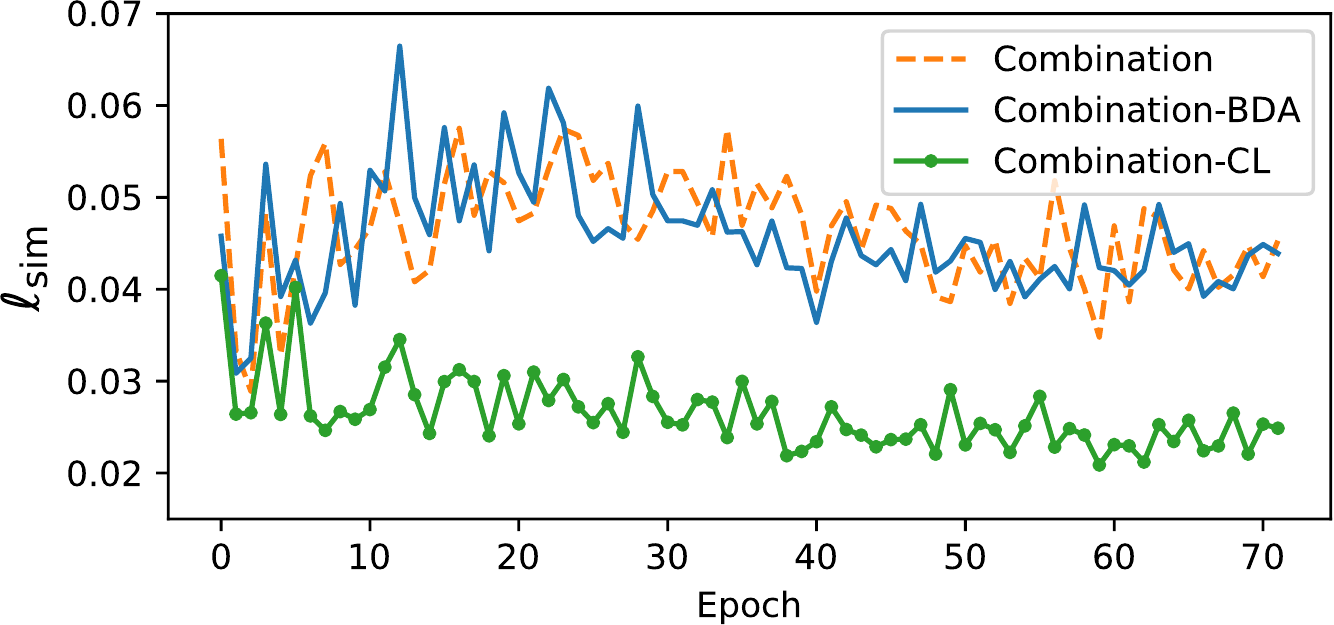}
    \caption{}
    \label{fig:js_loss_test}
  \end{subfigure}
  \caption{The consistency loss, \(\closs\), measured after each training epoch
           for the \textit{Combination} models on (a) the training set and (b)
           the test set of the fold 1 split.}
  \label{fig:js_loss}
\end{figure}

To confirm whether the consistency loss term is indeed enforcing consistency
more effectively than the cross-entropy loss on its own, we measured the
average JS divergence after each training epoch. This can be carried out for
any model provided the data is processed in triplets during the validation.
Figure \ref{fig:js_loss} plots the progress of the consistency loss term for
the training set and the test set of the fold 1 split -- specifically for the
\textit{Combination} models, albeit we observed similar patterns with the other
models. The figures show that the cross-entropy loss on its own encourages
consistency to some extent but not as effectively as having an explicit loss
term. It is interesting to note that using BDA resulted in a lower JS
divergence for the training set than standard data augmentation.

\section{Conclusion}
\label{sec:conclusion}

In this paper, we investigated consistency learning as a way to regularize the
latent space of deep neural networks with respect to input transformations
commonly used for data augmentation. We argued that enforcing consistency can
benefit tasks such as audio classification. We proposed using the
Jensen-Shannon divergence as a consistency loss term and used it to constrain
the neural network output for several audio transformations. Experiments on the
ESC-50 audio dataset demonstrated that our method can enhance existing data
augmentation methods for audio classification and that consistency is enforced
more effectively with an explicit loss term.

\vfill\pagebreak

\bibliographystyle{IEEEbib}
\bibliography{references}

\end{document}